\documentclass[11pt]{article}

\usepackage{natbib}
\setcitestyle{authoryear,open={(},close={)}}

  \usepackage[cmintegrals,cmbraces]{newtxmath}
\usepackage{ebgaramond-maths}
\usepackage[T1]{fontenc}
\usepackage{microtype}

\usepackage[top=.85in,left=1.5in,right=1.5in,bottom=.85in]{geometry}

\usepackage{setspace}

\usepackage{color,soul}

\usepackage{fancyhdr}
\usepackage{linguex}

\usepackage{titlesec}

\titleformat{\section}
  {\normalfont\fontsize{13}{13}\bfseries}{\thesection}{1em}{}

\titleformat{\subsection}
  {\normalfont\fontsize{12}{12}\bfseries}{\thesubsection}{1em}{}

\titleformat{\subsubsection}
  {\normalfont\fontsize{12}{12}\bfseries}{\thesubsubsection}{1em}{}

\usepackage[dvipsnames]{xcolor}

\newcommand{\bi}{\begin{itemize}}
\newcommand{\ei}{\end{itemize}}

\begin{document}

\setlength{\Extopsep}{.5em}
\setlength{\SubExleftmargin}{\parindent}
\setlength{\SubSubExleftmargin}{5em}

\noindent {\Large \bf{Algorithmic neutrality}} \vspace{.2in}

\noindent Milo Phillips-Brown\footnote{For many conversations and comments on drafts, thank you to Laura Alvarez Jubete, Serena Booth, David Boylan, Carol Brown, Thomas Byrne, medb corcoran, Ray Eitel-Porter, Sina Fazelpour, Nilanjan Das, Kevin Dorst, Dalia Gala, David Grant, Lyndal Grant, Sally Haslanger, Abby Jaques, Matt Mandelkern, Silvia Milano, Carina Prunkl, Bernhard Salow, Jen Semler, Kieran Setiya, Jack Spencer, Charlotte Unruh, Kate Vredenburgh, and audiences at  Northeastern University, the Massachusetts Institute of Technology, the Philosophy in an Inclusive Key Summer Institute, the Annual Meeting of the PPE Society, the University of Oxford, the University of Washington, and the University of Wisconsin. Thank you to Ginger Schultheis for extensive and transformative feedback. Thank you especially to Marion Boulicault, P. Quinn White, and the members of the Jain Family Institute's Digital Ethics and Governance team for years of inspiration, guidance, and encouragement. Work on this paper was funded by a gift from Accenture.}

\noindent University of Edinburgh \vspace{.15in}

\noindent forthcoming in \emph{Australasian Journal of Philosophy}\vspace{.15in}

\begin{quote}
\noindent \emph{Abstract.} Algorithms wield increasing power over our lives. They can and often do wield that power unfairly, and much has been said about \emph{algorithmic fairness}. In contrast, \emph{algorithmic neutrality} has been largely neglected. I investigate algorithmic neutrality, asking: What is it? Is it possible? And what is its normative significance? \vspace{.05in}
\end{quote}

\noindent In 2005, Adam and Shivaun Raff started a small business, Foundem, a compar\-is\-on-shop\-ping site similar to Google Shopping. Foundem showed promise, hailed at one point as the UK's best comparison-shopping site. But on June 26, 2006, Google altered its search algorithm; foundem.com fell from the top three search results to the 70s. By all indications, Foundem's demotion didn't reflect a decline in quality; foundem.com still sat near the top of Yahoo's and Microsoft's search results. But in the search engine optimisation industry, it's said that if you want to bury a body, you put it on the second page of Google. Foundem was no exception. It would not recover from the loss of traffic \citep{manthorpe2018}. 
 
The Raffs fought back, initiating the \emph{search neutrality} movement. They became the first plaintiffs in a case against Google that resulted in a \texteuro2.42 billion fine from the EU, which found that Google's search engine unfairly favoured Google's own product (Google Shopping) over competitors, including Foundem. 

More generally, \emph{algorithmic neutrality} is a central—and contested—notion in how we understand the digital world. The technology industry and the public often view algorithms as neutral \citep{benjamin2019}. Those who helm the world's most powerful algorithms portray neutrality as an ideal to strive for, one that they purport to meet:

\begin{quote}

I can assure you we [deliver search results] in a neutral way.\\
(Google CEO Sundar Pichai in \citep{cspan2018}, testifying before US Congress)

My goal is to be neutral. \\
(Meta CEO Mark Zuckerberg,\footnote{Zuckerberg here is talking both about his political donations and content moderation.} \citeyear{zuckerberg2024}, in a letter to the US House Judiciary Committee)

For Twitter to deserve public trust, it must be politically neutral.\\ 
(X CEO Elon Musk, \citeyear{musk2022})

\end{quote}

\noindent These are not isolated claims. Big Tech has long used rhetoric of neutrality as a shield against criticism \citep{guthrieweissman2017}; neutrality is a defining feature of how Google and Facebook see and present themselves \citep{gillespie2014}. (As interviews with fifty-one Facebook employees found, the ‘notion that Facebook is an open, neutral platform is almost like a religious tenet inside the company' \citep{thompson2018}.) {

Meanwhile, Big Tech's critics tell a different story:

\begin{quote}

There is no such thing as a morally neutral algorithm.\\
(Cathy O'Neill, \citeyear{oneill2017})

While we often think of terms such as `big data' and `algorithms' as being benign, neutral, or objective, they are anything but.\\
(Safiya Noble, \citeyear{noble2019}, p. 11)

Artificial intelligence is not… neutral.\\
(Kate Crawford, \citeyear{crawford2021}, p. 211)
\end{quote}

\noindent We find the same narrative far from Silicon Valley:

\begin{quote}
Artificial intelligence uses algebraic operations that are carried out in a logical sequence…. This method of calculation—the so-called ‘algorithm'—is neither objective nor neutral.\\
 (The late Pope Francis, \citeyear{francis2024})

\end{quote}

\noindent Politicians have entered the debate, alleging in high-profile hearings that Big Tech is not neutral.\footnote{See, for example, \citep{congress2018a,congress2018b}.} The courts, too, have turned their attention to algorithmic neutrality—when interpreting Section 230 of the US Communications Act (see \S\ref{generalising}), which sits at the heart of Internet regulation.

% https://ec.europa.eu/futurium/en/system/files/ged/platformneutrality_va.pdf

Despite all of this, the general notion of algorithmic neutrality has never been systematically studied. (In contrast, there is a mountain of literature about algorithmic \emph{fairness}, which I distinguish from neutrality in \S\ref{q3}. Two \emph{particular} kinds of algorithmic neutrality, search neutrality and net neutrality, have also received sustained scholarly attention.\footnote{For work on search neutrality see, for example, \citet{goldman2006, grimmelmann2010, crane2012, gillespie2014}.}) 

I investigate algorithmic neutrality, asking three questions. \emph{What is algorithmic neutrality?} I characterise neutrality in terms of the \emph{aim} of a given algorithmic system; an algorithmic system is neutral only if values other than that aim—values ‘external' to that aim—play no role in how the system delivers its results. (For example, if a search engine aims to give relevant results, then it is neutral only if values other than relevance—like the financial interests of the search engine operator or values of social justice—play no role in how the search engine delivers its results.) \emph{Is algorithmic neutrality possible?} The answer is \emph{no} for search engines and for most of the algorithmic systems that wield increasing power over our lives. \emph{What is the normative significance of algorithmic neutrality?} The full story defies a simple synopsis, but the headline is that neutrality is a red herring. Algorithmic neutrality—in and of itself—is neither good nor bad, neither fair nor unfair. It is not—in and of itself—something to strive for or avoid. The true normative questions concern what roles values play in how algorithms deliver their results. Algorithmic neutrality is, then, a false ideal and a faulty shield.

In answering the three questions, I'll work first and mainly with a case study, search engines, and then consider algorithmic systems of all kinds. 

%% GENAI

%%%%% WHAT IS SEARCH NEUTRALITY?
%%%%% 
%%%%% 

\section{What is search neutrality?}\label{q1}

To answer this question, it will help to characterise a general kind of neutrality---algorithmic or otherwise---of which I'll argue search neutrality is an instance.
 
Imagine a scientist who finds evidence that a pandemic-causing virus originated in a lab of a major geopolitical power; publishing this evidence, she thinks, would spark global conflict, and so she discards it. Another scientist, funded by a tobacco company, disregards signs that smoking causes cancer. Neither scientist conducts her research neutrally. Or imagine that you are a tour guide, charged with showing tourists local attractions. Instead, you often take tourists to stores that give you a kickback for every customer you bring them. You do not perform your job neutrally. Finally, consider that many contemporary copy machines will not copy paper money: if you give them paper money and hit \emph{copy}, nothing will happen. Such copy machines don't perform their function—the function of copying—neutrally. 

To articulate the general kind of neutrality common to these cases, I'll begin with a particular, well-studied kind of neutrality—neutrality in science\footnote{Or, as it's often put, \emph{value-freedom} in science. Others—for example, \citet{dotan2020}, \citet{fazelpourdanks2021}, and \citet{johnson2023}—also connect this literature to questions about algorithms, as I discuss in \S\ref{general-q2}.}—that is standardly defined along these lines:\footnote{See, for example, \citep{douglas2009}.}

\ex.[] \emph{Neutrality in science} \\ Science is neutral only if non-epistemic values play no role in how scientists conduct core scientific practices.

\noindent For our purposes, we need only a basic understanding of the terms in this definition. First, what distinguishes epistemic from non-epistemic values? Pursuing epistemic values—like empirical adequacy or internal consistency—leads us to truth \citep{steele2012}. (It doesn't matter here just what empirical adequacy and internal consistency are.) Pursuing non-epistemic values—like financial interests and political ideologies—doesn't lead us to the truth. Second, what is a core scientific practice? The idea is that neutrality is compromised if non-epistemic values play a role in how scientists engage in certain scientific practices (like gathering evidence), but not in others (like applying for grants). The former are core scientific practices; the latter aren't. 

Putting these pieces together in our examples: science isn't neutral in the case of our two scientists because a non-epistemic value (world peace, funders' financial interests) plays a role in how the scientists conduct a core scientific practice (gathering evidence).

There are many questions you could ask about scientific neutrality: How to distinguish epistemic and non-epistemic values? What counts as a core scientific practice? Can science be neutral, and if so, should it be? These are not our questions. What matters here is the kind of neutrality at stake: why should scientific neutrality require that non-epistemic values play no role in core scientific practices? I suggest that it's because science \emph{aims at truth} \citep{reisssprenger2020}. Truth is science's north star. We might say that epistemic values are \emph{internal} to this aim, while non-epistemic values are \emph{external}. If values that don't lead us to the truth—external values—play a role in core scientific practices, those practices deviate from science's aim. Science is thereby not neutral. 

 This suggests a general kind of neutrality---in terms of the aim of a system or practice and values that are either internal or external to that aim. Your role as a tour guide aims to show tourists local attractions. When values external to this aim—your financial interests—play a role in how you guide tours, your practice is thereby not neutral. The aim of a copy machine is to make copies. When values external to this aim—the integrity of the monetary system—play a role in how the machine makes (or doesn't make) copies, its ‘practice' is thereby not neutral.

We, then, can characterise search neutrality in terms of the aim of search engines. What is that aim? The standard answer is \emph{relevance} \citep{munton2022}, a notion I'll explore in \S\ref{q2}, as we see in, for example:

\begin{quote}
Search engines do perform the function of guiding the enquirer to sources of testimony which are supposed to be relevant.\\ (Simpson 2012, p. 429, emphasis mine)

The function of search engines is to literally provide a testimony to the user about what information is available and relevant to her query.\\ \citep[p. 234]{elgesem2008}

Search algorithms have a set of organizing criteria for the kind of phenomena they seek: particular kinds of websites, particular patterns of incoming links, and particular behaviors of users, all read as signals of a genuinely emergent and non-strategic demonstration of a site's true relevance.\\ \citep[p. 65]{gillespie2014}

\end{quote}

\noindent While these authors don't use the term \emph{aim}, they share the idea that what search engines \emph{do}—what their function is, what their organising criteria are—is give relevant results. When I say that search engines aim at relevance, I have this shared idea in mind. (If you prefer a term other than \emph{aim}, feel free to use it.) 

 And so I propose:

\ex.[] \emph{Neutrality in search}\label{relevance-neutrality} \\ A search engine is neutral only if values other than relevance play no role in how the search engine ranks pages.

 \noindent Relevance is internal to the aim of search engines; all other values are external. 

Imagine a search engine operator that intentionally ranks pages of its own products above those of its competitors' even when its competitors' pages are more relevant. (This doesn't take much imagining; it's exactly what the EU fined Google for doing.) A value other than relevance---the external value of the operator's financial interests---plays a role in how the search engine ranks pages. The search engine is thereby not neutral. Or imagine a search engine that, when given the query \emph{origins of pandemic virus}, suppresses results that suggest that the virus originated in a lab. A value other than relevance---the external value of averting global conflict---plays a role in how the search engine ranks pages. The search engine is thereby not neutral. (Note: I'll be discussing search engines that return a list of pages. What I'll say also applies to generative-AI based search engines that instead directly answer the query.)

Compare how I've characterised search neutrality with how it's normally characterised by search engine operators, their critics, and academics alike:

\begin{quote}

We [i.e. Google] do get concerns across both sides of the aisle. I can assure you we do this [i.e. deliver search results] in a neutral way. And we do this based on a specific keyword, what we are able to assess as the most relevant information.\\
 (Google CEO Sundar Pichai in \citep{cspan2018})

Search Neutrality can be defined as the principle that search engines… \ should have no editorial policies other than that their results be… \ based solely on relevance.\\
 \citep{searchneutrality2009}

Search neutrality... at its heart is some idea that Internet search engines ought... [to] employ ‘neutral' search algorithms that determine search result rankings based on some ‘objective' metric of relevance. \\
\citep[p. 1199]{crane2012}

\end{quote}

\noindent It's no accident that these characterisations coincide with mine. Implicit in them is what I've made explicit: when we take relevance as the aim of search engines, we can---implementing the general notion of neutrality that's based on aims---characterise search neutrality in terms of relevance.

\subsection{Clarification: what's in an aim?}\label{aims}

I noted above that it's the standard view that search engines aim at relevance. The sense of \emph{aim} at issue must be understood in a particular way, since there is another sense on which search engines need not simply aim at relevance. Consider, for example, a search engine that doesn't rank pages on the basis of relevance when doing so wouldn't serve its financial interests. In some sense of \emph{aim}, such a search engine does not aim simply at relevance. 

What sense of \emph{aim}, then, does the standard view concern? Jessie Munton offers a compelling answer:

\begin{quote}
[The fact that] all major search engines are run by profit-making companies… \ skews their design away from the generation of maximally relevant [results]… \ Their goal, in practice, is \emph{not} to simply return information… [But] it is not, at a conceptual level, \emph{part of the task of a search engine} to generate profit. That is a contingent feature of their actual operation.\\ 
\citep[p. 7, original emphasis]{munton2022}
\end{quote}

\noindent The sense in which all search engines aim at relevance, Munton tells us, is \emph{conceptual}. (We could also call it \emph{constitutive} or \emph{idealised}—again, use whatever label you like.) The other sense is \emph{contingent}.

Just what is the conceptual sense? I don't have space to do this question justice—there's an entire literature on the related issue of the nature of \emph{functions} in artifacts generally. The notion of a conceptual aim is nonetheless intuitive, and we can bring it into sharper focus by considering how it arises in other cases.

We can, for example, distinguish conceptual and contingent aims with the scientists who discard evidence in pursuit of non-epistemic values (preventing world war, serving funders' financial interests). The scientists' practices, as a matter of fact, don't aim just towards the truth, but also towards furthering other values—these are their contingent aims. But nonetheless it's commonly accepted that scientific practices aim at the truth—this is the conceptual aim of all scientists' practices. Or consider that a copy machine that will not copy paper money doesn't, as a matter of fact, simply aim at copying. The copy machine is nonetheless \emph{a copy machine}—its conceptual aim is copying.

Let me emphasise again that the issue here is not particular to my project. Anyone who subscribes to the standard view, that search engines aim at relevance, must understand that view in this conceptual sense. What is particular—and novel—to my project is characterising neutrality in terms of conceptual aims. (In what follows, I'll shorten \emph{conceptual aim} to \emph{aim}.) 

%%%%% IS SEARCH NEUTRALITY POSSIBLE?
%%%%% 
%%%%% 

\section{Is search neutrality possible?}\label{q2}

Search neutrality is impossible. To see why, let's begin with the notion of a \emph{multidimensional} concept. Take basketball skill, for example. Jack is a better dribbler than Nashid, but a worse shooter. Is Jack a more skilled basketball player? There may be no good answer to this question. Yes, along one \emph{dimension} of basketball skill, dribbling ability, he is more skilled. Along another dimension, shooting ability, he isn't. But is he more skilled \emph{overall?} To answer this question, we need some way to compare Jack's dribbling to Nashid's shooting. We need to weigh the dimensions of basketball skill against one another. 

How should they be weighted? Many theorists maintain that there is no privileged weighting for multidimensional concepts \citep{kamp1975, sen1997, parfit2016}, concluding that multidimensional concepts therefore generate \emph{incomparability}. Jack is neither more nor less skilled than Nashid, nor are they equally skilled. They are incomparable in basketball skill. 

The thesis that multidimensionality yields incomparability is intuitive and widely accepted, although some contest it (for example, \citet{dorretal2023}). A strong version of the thesis says that whenever there's multidimensionality, there is incomparability. A weaker version, which I adopt, says that this is only sometimes so. For now, I'll assume that multidimensionality yields incomparability; in \S\ref{purposes} I'll consider when it might not.

%% CHANGE TO SOMETHING MORE INTERESTING IF ACCEPTED

Relevance, I maintain, is a multidimensional concept. Imagine that you enter \emph{hurricane} into a search engine. What is relevant to the query? There are many candidates: what a hurricane is; whether global warming supercharges hurricanes; the human toll of hurricanes; whether that toll differs across racial lines; natural disasters similar to hurricanes, like tsunamis or tornadoes; hurricanes that are often discussed (for example, Hurricane Katrina); etc.

Now consider a toy case of two webpages. The first is detailed, discussing how hurricanes form, Hurricane Hortense (an unremarkable hurricane), and the nature of forest fires. The second is brief, discussing Hurricane Katrina and the nature of tornadoes. The first page is more relevant than the second along some dimensions of relevance but not along others.

What are these dimensions? One is \emph{aboutness}. Facts about how hurricanes form, for example, are about hurricanes in a way that facts about, say, the nutritional value of mushrooms aref not. Another is \emph{salience}. Hurricane Katrina is more salient than Hurricane Hortense, and so the page that discusses Katrina is more relevant, along the dimension of salience, than the one that discusses Hortense. Another dimension is \emph{similarity}; the page that discusses tornadoes is more relevant along this dimension than the page that discusses forest fires. Another---and there are no doubt more---is \emph{informativeness} \citep{roberts2012}: the amount of information concerning topics that are relevant along the other di\-men\-sions—topics that are about the search term, or that are salient, or that are similar. The first, more detailed page is more relevant along this dimension than the second, brief one.

Is the first page more, less, or equally relevant as the second \emph{overall}? To answer, we'd need a privileged weighting of the dimensions, but there is none, and so the two pages are incomparable in relevance.

We can now give a first-pass statement of my argument. A search engine that aims to rank pages simply on the basis of relevance cannot rank one of these pages above the other. The best it can do is to rank one above the other on the basis of a given weighting of the dimensions of relevance. But which? One can't appeal to the aim of relevance itself to answer this question because there is no privileged weighting. In other words, the aim of relevance \emph{underdetermines} which weighting to use in ranking pages, and so underdetermines how to rank pages. Values other than relevance must play a role in weighing these dimensions, and so must play a role in ranking pages. Search neutrality is impossible.\footnote{\label{the-others}Others also argue that neutral search is somehow impossible. Most prominently, Grimmelmann (2010) objects to various ways of characterising search neutrality—alleging that some are incoherent, others impossible, and still others lacking for other reasons. His allegations of \emph{impossibility} do not concern the (standard) characterisation of neutrality I adopt. Rather, his objection to this characterisation is that it's unworkable as a basis for \emph{policy}. Since policy is not my concern, I leave it to future work to explore how my discussion relates to Grimmelmann's.} (My argument parallels prominent arguments, like Rudner's (\citeyear{rudner1953}), that scientific neutrality is impossible on the grounds that the aim of truth underdetermines how to conduct core scientific practices.)

\subsection{Objection: the searcher's purposes}\label{purposes}

An objector might question whether relevance underdetermines how to weigh the dimensions of relevance. Her reasoning starts with an observation I agree with: using search engines is a way of inquiring, and in general, what's relevant to an inquiry varies with its \emph{purpose}, as \citet{garfinkel1981} evocatively illustrates:

 \begin{quote}

When [infamous bank robber] Willie Sutton was in prison, a priest who was trying to reform him asked him why he robbed banks. ‘Well,' Sutton replied, ‘that's where the money is.' Clearly there are different… \ purposes shaping the question and answer. [The priest and Sutton] take different things... to stand in need of explanation. (p. 21)

\end{quote}

In the case of search, what's relevant to a query varies with the purpose of the searcher. Imagine one person with a purpose like Sutton's, and another with a purpose like the priest's. Each enters \emph{Why did Sutton rob banks?} into a search engine. For the first person, the most relevant pages concern issues like the financial windfall, degree of difficulty, and risk of imprisonment in robbing banks. For the second, the most relevant pages concern issues like Sutton's motives, character, or religious background. 

So, we shouldn't talk \emph{simply} of whether a page is more relevant than another, but instead this talk should be indexed to a given purpose. This is why my argument, as I stated it earlier, was a first pass. I concluded simply that the aim of relevance underdetermines how to weigh the dimensions of relevance. This conclusion, properly stated, is that \emph{for some purposes}, the aim of relevance underdetermines how to weigh the dimensions of relevance. My objector and I agree about this.

But the objector goes further, maintaining that the searcher's purposes always and fully determine these weightings, and so the multidimensionality of relevance never yields incomparability after all.

I can (and do) happily grant that some purposes fully determine these weightings. But I deny that all do. Imagine a child who searches \emph{hurricanes} with the purpose of exploring the topic of hurricanes. She is simply curious to learn about them; she has no particular aspects or facts about hurricanes in mind. Her \emph{exploratory purpose}---and exploratory purposes more generally---doesn't determine how to weigh the dimensions of relevance. (I suspect, but don't need for my argument, that the purposes of many searches are partly exploratory.) Search neutrality remains impossible.

\subsection{Objection: randomness}\label{randomness}

Another objector might grant that relevance underdetermines how to weigh the dimensions of relevance, but resist my conclusion that search neutrality is therefore impossible. This objector says that if one picks \emph{randomly} among the possible weightings of the dimensions of relevance, then neutrality is possible. After all, it's natural to think in general, one is neutral if one picks randomly.

The objection fails. Consider an analogy: there's an upcoming presidential election with two Democratic candidates and four Republicans; a certain newspaper can pick only one candidate interview and it commits to doing so randomly; but randomly among what? It could pick randomly among individual candidates, giving each candidate an equal (one-in-six) chance. Or it could pick randomly along party lines, giving each party an equal (one-in-two) chance. Or it could pick randomly along any number of lines: gender lines, racial lines, etc.

The objector must say that if the newspaper picks randomly among individual candidates, it's neutral among individual candidates—but \emph{not} neutral along party lines. (There is a two-in-six chance that a Democrat is interviewed, but a four-in-six chance that a Republican is interviewed.) If the newspaper picks randomly along party lines, it's neutral along party lines—but not among individual candidates. (Each Democrat has a one-in-four chance of being interviewed, but each Republican has a one-in-eight chance.\footnote{I'm assuming, for simplicity, that both Democrats have the same chance as the other and every Republican has the same chance as the others.}) So, in picking candidates to interview, the newspaper can be neutral among individual candidates or it can be neutral along party lines, but it can't be neutral in both ways at once.

The same goes for search engines. Consider the search \emph{interview of presidential candidates} in a case where each candidate has been interviewed by a different newspaper, and each newspaper has a webpage for its own interview. The pages are, imagine, incomparable with respect to relevance. The search engine can be neutral among individual candidates or it can be neutral along party lines, but it can't be neutral in both ways at once. Search neutrality remains impossible.

\subsection{Clarification: the scope of the impossibility}\label{clarification} 

In discussing the normative significance of search neutrality (in \S\ref{q3}), it will help to understand the scope of the impossibility. This requires distinguishing two roles external values can play in ranking pages: they can \emph{override} relevance or \emph{supplement} it.

I've argued that because relevance is multidimensional, pages are sometimes incomparable in relevance (for some purposes). But multidimensionality does not always breed incomparability. Recall Jack and Nashid: Jack is a more skilled basketball player along one dimension, but not another. They are incomparable in basketball skill \emph{overall}. There are nonetheless cases of comparability in basketball skill: for example, LeBron James simply \emph{is} a more skilled basketball player than I am. He is a more skilled player than I am along all dimensions.

Similarly, while some pages are incomparable in relevance, other pages are comparable: some pages are more relevant than others, overall. If you enter \emph{LeBron James} into a search engine, pages that discuss LeBron James are more relevant, overall, than my professional website. The former pages are more relevant than the latter along all dimensions. (It seems likely, as noted in \S\ref{purposes}, that comparability sometimes arises even when one page is more relevant than another only along some dimensions.)

External values \emph{override} relevance in cases of comparability. In particular, an external value overrides relevance just if a search engine ranks a less relevant page above a more relevant one (or vice versa) because that external value is at play. For example, when the EU fined Google, it alleged that Google, in pursuit of financial gain, ranked pages of its own products above those of its competitors even when the former were less relevant than the latter: Google's financial interests overrode relevance.
% citation \footnote{{\blue Cite this: https://ec.europa.eu/commission/presscorner/detail/ro/memo\_17\_1785.}}

External values \emph{supplement} relevance in cases of incomparability in relevance. In particular, an external value supplements relevance just if, given two pages that are incomparable in relevance, the search engine ranks one above the other because an external value is at play. Say that we have two pages that are incomparable in relevance; ranking the first above the second would better serve the financial interests of a search engine operator; and for this reason, the search engine is designed to rank the first above the second in any such scenario. These financial interests supplement relevance. 

We can now see the scope of the impossibility. Search neutrality, we know, is impossible because external values must play some role in how a search engine ranks pages \emph{in cases of incomparability}. Stated in our new terminology: external values necessarily supplement relevance. It's not necessary, though, that external values override relevance. For example, Google could simply not rank pages of its products above other, more relevant pages.

\section{What is the normative significance of search neutrality?}\label{q3}

There are many things to say about algorithmic neutrality's normative significance. Here, I focus on three aspects.

\subsection{Neither good nor bad, neither fair nor unfair}\label{lessons-1}

As I noted in the introduction, algorithmic neutrality is often viewed as a goal \emph{in and of itself}, and non-neutrality as something to avoid. Search neutrality in particular has been held up by Google as ideal.

Neutrality is, in fact, a false ideal. Search engines may be non-neutral in different ways; some of these ways are good or fair and others not. Similarly, were a neutral search engine possible, it would in some respects be good or fair, and in other respects not. What matters normatively is not whether a search engine should be neutral \emph{per se}. Whether a search engine is neutral is a matter of whether external values play a role in how a search engine ranks pages. And so what matters normatively is which external values should play a role in how search engine ranks pages, and what role they should play. We see this both with external values overriding relevance and supplementing it.

Consider, for example, a search engine that will not return results that contain child pornography. Since child pornography is relevant, overall, to the purposes of searchers interested in finding child pornography, the search engine overrides relevance with external values—values of protecting children. What's the normative status of this overriding non-neutrality? Presumably, it's good. And that's not because non-neutrality is good in and of itself. Rather, it's because it's good to protect children, even at the ‘cost' of not delivering relevant results. What matters, normatively, is whether a certain external value (protecting children) should play a certain role (overriding relevance) in how a search engine ranks pages.

Other times, non-neutrality from overriding relevance is bad. It's bad when Google overrides relevance to serve its own financial interests (or at least the EU saw it this way).  That's not because non-neutrality is bad in and of itself, but rather because this particular form of non-neutrality is unfair. What matters, normatively, is whether a certain external value (that of Google's financial interests) should play a certain role (overriding relevance) in how a search engine ranks pages.

We can bring out the point further with Grimmelmann's discussion of the query \emph{jew}:

\begin{quote}
Google has been criticised both for returning antisemitic sites (to American users) and for \emph{not} returning such sites (to German users). The inescapable issue is that Google has users who want to read anti-Semitic web pages and users who don't. … \ helping users find what they want is such a profound social good that one should be sceptical of trying to inhibit it. Telling users what they \emph{should} see is a serious intrusion of personal autonomy… \ If you want Google to steer users to websites with views that differ from their own, your goal is not… \ \emph{neutrality}.\\ \citep[pp. 446--7, original emphasis]{grimmelmann2010}
\end{quote}  

\noindent Anti-Semitic sites are relevant, overall, to the purposes of some searchers. The aim of relevance  dictates that a search engine would show such sites to such users. To do otherwise would, as Grimmelmann notes, violate neutrality. External values—maybe values of social justice—would play a role in how the search engine ranks pages. They would override relevance. 

What is the normative significance of this situation? Grimmelmann suggests it's an unhappy one; the German government disagrees. My goal in this paper isn't to adjudicate questions like these; that would require investigating how search engines fit within particular social and political context. I aim to characterise the \emph{structure} of such questions. And that structure doesn't concern neutrality \emph{per se}: what's at issue is whether certain values (those of social justice) play a certain role (overriding relevance, and so compromising autonomy.)

The question of whether certain external values should override relevance can be put another way: to what extent is relevance worth pursuing? Is relevance worth pursuing at the price of children being abused? No. Is relevance worth pursuing at the ‘price' of compromising Google's financial interests? Yes. Is relevance, and its boon to autonomy, worth pursuing at the price of compromising values of social justice? I leave that to others to work out.

We saw in \S\ref{clarification} that it's not inevitable that external values override relevance. And so the normative question is \emph{whether} external values should override relevance, and if so, which. But it is inevitable that external values supplement relevance. The question is then not whether external values should supplement relevance (they can't do otherwise), but rather \emph{which} should. For any given value, we ask whether it should be among those that supplement relevance. I'll consider two issues that bear on this question now, and then a third in \S\ref{lessons-2} . 

The first issue arises in cases of \emph{nudging}. In cases of nudging, some entity or agent intervenes on the options for action a subject might take. Which options are available to the agent remain the same, but they're presented in a different way, one that nudges the agent to take certain options rather than others. If a grocery store carries both healthy and unhealthy cereals, but places healthy cereals at a customer's eye level, her options are not changed (both kinds of cereal are available), but they're presented to her in a way that nudges her towards picking the healthy option.

Consider that sometimes, when people who search \emph{suicide}, certain sites that discuss how to commit suicide are incomparable in relevance with sites that dissuade people from committing suicide. The search engine operator ranks the latter pages above the former; in this case, the value of preservation of life supplements relevance. The searcher is nudged towards anti-suicide sites. What's the normative significance of this? It is: should a certain external value (preservation of life) play a certain role (supplementing relevance)? Put another way: should the value of preservation of life be among the external values that supplement relevance? 

The second issue arises in cases parallel to ones involving \emph{positive action}. Positive action, a notion in  United Kingdom law, is one kind of affirmative action involved in United States law. Consider a case of hiring. You have two, equally qualified applicants, one black and another white (more generally, one from a disadvantaged, protected group and another not). You take positive action if you favour the black applicant (by, for example, hiring them) because the black applicant is black and the white applicant isn't. 

Imagine that whenever sites from a black-owned business are tied or incomparable in relevance with sites from white-owned businesses, a search engine treats the former site more favourably (by ranking higher) because the former but not the latter is black-owned. Values of social justice supplement relevance. When we ask about the normative status of this case, we ask, again, whether certain values of social justice should be among those that supplement relevance. 

Having surveyed these cases, I can deliver on a promise I made earlier—to explain how algorithmic neutrality relates to algorithmic fairness. We've encountered cases of non-neutrality without fairness, as in the case of Google's fine from the EU, where Google unfairly wielded its monopoly in overriding relevance with its financial interests. 

We can imagine a similar case of non-neutrality \emph{with} fairness. Google's practices are unfair partly because Google dominates the market: when Google's financial interests override relevance, its would-be competitors can't compete, and that's unfair. But imagine a different world where the market is split among forty search engines, each with less than 3\% market share. Some search engines override relevance with their business interests (and the operator tells users that it's doing so), showing pages that sell their own products over more relevant ones. What is the normative significance of this situation? We might suppose, inspired by \citet{goldman2006}, that it's fair for these search engines to override relevance. They are a part of a non-monopolistic market.\footnote{We can distinguish two questions here. One is whether the \emph{market} is fair: in the non-monopolistic, hypothetical case, the market is fair; in the actual, monopolistic case, the market is not fair. My concern is instead with the related question of whether the search engines themselves are behaving fairly: in the non-monopolistic, hypothetical case, the search engines that override relevance with their own financial interests behave fairly; in the actual, monopolistic case, Google's search engine doesn't behave fairly.} Search engine operators may fairly override relevance with their financial interests; searchers can decide whether to use such search engines.

We've already encountered other cases of non-neutrality with fairness: practices analogous to positive action are non-neutral but are, presumably, sometimes fair. Relatedly, we can imagine cases of neutrality—were such a thing possible—without fairness. A neutral search engine doesn't engage in practices analogous to positive action; when such practices are fair, the neutral search engine is unfair. 

The normative significance of (non)neutrality extends beyond (un)fairness. For example, we've seen how a non-neutral search engine may objectionably impinge on autonomy or laudably protect children.

\subsection{Global and local neutrality}\label{lessons-2}

In the introduction, I noted the search neutrality movement's demands for search neutrality and Google's avowals of it. Demands and avowals of neutrality presuppose that neutrality is possible, but in fact it isn't. So how should we understand them?

The kind of neutrality I've characterised is \emph{global}: it requires that external values \emph{of any kind} play no role in how the search engine ranks pages. Both the search neutrality movement and Google understand search neutrality in the global sense, as we see in the quotes from \S\ref{q1}, repeated here:

\begin{quote}

We [i.e. Google] do get concerns across both sides of the aisle. I can assure you we do this [i.e. deliver search results] in a neutral way. And we do this based on a specific keyword, what we are able to assess as the most relevant information. \\
(Google CEO Sundar Pichai in \citep{cspan2018})

Search Neutrality can be defined as the principle that search engines… \ should have no editorial policies other than that their results be… \ based solely on relevance.\\
 \citep{searchneutrality2009}

\end{quote}

\noindent We know that Google's claim to global neutrality is not true because it cannot be true, and the search neutrality movement's demand for global neutrality is inapt because it cannot be met.

Where do we go from here? One option is to simply stop avowing and demanding neutrality. Another option is to retreat to a different, \emph{local} kind of neutrality. While global neutrality prohibits external values of any kind from determining search results, the local neutrality I have in mind prohibits only values \emph{of a certain kind}. For example, a search engine is locally neutral with respect to political values only if political values play no role in how the search engine ranks pages. 

Google could then avow this local, political neutrality when faced with accusations of political favour\-itism. The search neutrality movement could demand local, financial neutrality, which prohibits the financial interests of a search engine operator from playing a role in how the search engine ranks pages. 

Both of these local kinds of neutrality are possible; there is no conceptual trouble in demanding or avowing them. The story does not end there, though. Retreating to local neutrality raises new normative considerations.

Imagine a search engine—of the kind that Google claims to be—with which political values play no role in how  pages are ranked: the search engine operator doesn't attempt to advantage any particular political parties, figures, policies, ideologies, etc. Imagine further that the search engine's operators heavily weigh salience—which, recall, is a dimension of relevance—in designing their search engine. This choice will tend to generate rankings with higher spots for pages about certain political parties (and figures, etc.): those that are more salient. 

Now suppose that some minority political party—the P Party—struggles for public awareness. Heavily weighing salience disadvantages the P Party. Pages discussing the P Party will tend to be ranked lower, thereby compounding its struggles for public awareness. The P Party would, then, have certain standing to complain: the search engine operator could just have well not disadvantaged the P Party, since it could have weighted salience less heavily.\footnote{For a related issue, see Patten's (2011) distinction, in discussing liberal neutrality, between what he calls ‘neutrality of aim' and ‘neutrality of effects'.}

For contrast, consider a similar complaint that lacks this standing. Imagine that pages that discuss a different political party—the Q Party—tend to be less relevant than those that discuss rival parties \emph{no matter how the dimensions of relevance are weighted}. A search engine ranks pages discussing the Q Party lower—that is, ranks them according to their relevance—which disadvantages the Q Party. Despite this disadvantage, the Q Party lacks the P Party's standing to complain. Insofar as the search engine aims to give relevant results, the Q Party's disadvantage is ‘deserved'. Pages discussing them simply tend to be less relevant. In contrast, the P Party's disadvantage isn't deserved, since pages discussing it aren't simply less relevant. 

The issue generalises beyond the political domain. Consider that small businesses (like Foundem) are often less salient than bigger ones (like Google). The lower salience is weighted, the lower smaller businesses will be ranked (and in turn get fewer customers). Even if the financial interests of the search engine operator play no role in how the search engine ranks pages, small businesses could nonetheless be disadvantaged in a way that gives them  standing to complain. 

The issue also generalises beyond salience. Relevance has other dimensions, and different kinds of pages will tend to be ranked higher (or lower) depending on how the dimensions are weighted. These differences in rankings engender the kind of legitimate complaints like those available to the P Party or small businesses in the case of weighting salience.

What does this all mean for demands and avowals of neutrality from Google and the search neutrality movement? 

Google's avowal of neutrality is supposed to reply to charges of political fav\-our\-itism. The global version of this claim is false by conceptual necessity. If the local version were true, however, then the reply would partly succeed: local neutrality means that Google hasn't attempted to favour one political party over another. 

But it would succeed only partly. Even if political values play no role in how the search engine ranks pages, the weighting of relevance's dimensions might advantage certain parties, figures, policies, or ideologies, etc. This gives the disadvantaged party (figure, etc.) standing to complain of a certain kind of political favouritism.

To fully dispel concerns of political favouritism, then, Google would have to do the messy work of showing why its weighting of the dimensions of relevance doesn't confer advantages along political lines. I suspect that every weighting will confer some political advantage or other (partly for reasons I discussed in \S\ref{randomness}), but it's for future work to substantiate that suspicion.

For the search neutrality movement, there is a question of whether to demand more than just the local neutrality with respect to search engine operators' financial interests. They might demand further that search engine operators not weigh the dimensions of relevance in a way that advantages their financial interests, regardless of any intent to advantage themselves.

\subsection{Search bias}\label{lessons-3}

The impossibility of search neutrality threatens to rob complaints of search \emph{bias} of their normative force. After all, if no search engine is neutral, then isn't every search engine in some sense biased? And as Antony (\citeyear{antony1993})\label{antony} provocatively asks when discussing bias in epistemology: ‘If bias is ubiquitous and ineliminable… \ what are we complaining about?' (p. 136). Such scepticism towards search bias is indeed warranted sometimes, but often it's unwarranted—mainly for reasons we've already seen.

Recall our statement of search neutrality: 

\ex.[] \emph{Neutrality in search} (repeated from p. \pageref{relevance-neutrality})\\ A search engine is neutral only if values other than relevance play no role in how the search engine ranks pages.

\noindent I've characterised roles that external values may play in how a search engine ranks pages—overriding relevance or supplementing it. There are two corresponding kinds of bias:

\ex.[] \emph{Overriding bias} \\ A search engine is overriding biased if values other than relevance override relevance in how the search engine ranks pages.

\ex.[] \emph{Supplementing bias} \\ A search engine is supplementing biased if values other than relevance supplement relevance in how the search engine ranks pages.

\par There is no reason at all for scepticism about overriding bias. It's not inevitable that external values override relevance, as we saw in \S\ref{clarification}. (There are other forms of search bias that are also not inevitable. \citet{noble2019} exposed search engines for delivering disturbingly sexualised results for the query \emph{Black girls}—even to searchers who were not looking for sexualised material. This is a form of bias, and it's not inevitable. Search engines could simply not return such results. Similarly, Introna and Nissenbaum (2000) showed how the technical architecture of search engines of their day unfairly privileged sites of the wealthy and powerful. This is a form of bias that is, again, not inevitable. Introna and Nissenbaum showed that other architectures wouldn't have this effect.)

Scepticism \emph{is} warranted towards certain complaints of supplementing bias, but not all. If one complains \emph{that} external values supplement relevance, then one's complaint should be dismissed, since—as we know from \S\ref{lessons-1}—it is inevitable that external values supplement relevance. But complaints about \emph{which} external values supplement relevance don't warrant scepticism because it's not inevitable that certain values rather than others supplement relevance. And there are various reasons—as we know from \S\ref{lessons-1} and \S\ref{lessons-2}—for thinking that certain values rather than others should supplement relevance.

With all of the above considerations in hand, we can respond to a prominent sceptic about search bias, Gillespie (2014), who writes:

\begin{quote}

To accuse [a search] algorithm of bias implies that there exists an un-biased judgment of relevance available, to which the tool is failing to hew. Since no such measure is available, disputes over algorithmic evaluations have no solid ground to fall back on. (Gillespie, 2014, p. 175)

\end{quote}

\noindent Gillespie doesn't say what he means by ‘un-biased judgments', so I hesitate to respond with too much confidence. But in many cases, un-biased judgments seem clearly available: sites that discuss LeBron James are more relevant to the query \emph{LeBron James} than my academic website; antisemitic sites are relevant to searches of \emph{jew} for searchers whose purposes concern antisemitism; etc. Maybe in cases of incomparability, there is a sense in which un-biased judgments are unavailable. But as we've just seen, complaints of certain forms of search bias can nonetheless be apt in such cases.\footnote{Maybe Gillespie has other forms of search bias in mind, about which complaints of bias indeed warrant scepticism.}

\section{Algorithmic neutrality in general}\label{generalising}

What is algorithmic neutrality? Is it possible? And what is its normative significance? These are the questions that animate this paper. I'll now answer them, generalising my answers about search engines.

\subsection{What is algorithmic neutrality?}\label{general-q1}

I argued that there's a general kind of neutrality—algorithmic or otherwise—that we can characterise in terms of the aim of a given system or practice. When applied to search engines, we get that a search engine is neutral only if values other than relevance—values external to its aim—play no role in how it ranks pages. More generally, an algorithmic system is neutral only if values external to its aim play no role in how it delivers its results.

For example, imagine a hiring algorithm that aims to rank candidates on the basis of qualification. The algorithm is neutral only if values other than qualification play no role in how the algorithm ranks candidates. So, it's not neutral if an applicant's politics, race, or family connections play any role in how it ranks candidates. In such cases, values external to qualification play a role in its rankings.

As just noted, I motivated my view about search neutrality with general considerations about the nature of neutrality. And, as I discussed in \S\ref{q1}, my view of search neutrality coincides with how it's commonly understood; implicit within this common understanding, I said, is a view of neutrality based in a search engine's aim. We find this view implicit in other characterisations of algorithmic neutrality.

Consider the oral arguments in the US Supreme Court of \citet{gonzalezvgoogle2023}, which concerned, among other things, the thumbnails displayed by You\-Tube's recommendation algorithm, and whether they in some way advantaged videos that favour ISIS. Algorithmic neutrality was a theme throughout, as a few excerpts illustrate:

\begin{quote}
[Justice Thomas:] the thumbnails… [are] \emph{based upon what the algorithm suggests the user is interested in}… it's neutral in that sense… Say you get interested in rice—in pilaf from Uzbekistan. You don't want pilaf from some other place, say, Louisiana… I don't see how that is any different from what is happening in this case… are we talking about \emph{the neutral application of an algorithm that works generically for pilaf and also works in a similar way for ISIS videos?} (emphasis mine)
\end{quote}

\noindent Here we see the assumption that the recommendation system aims at meeting users' interests: the recommendations are ‘based upon what… the user is interested in'. We also see the idea that the algorithm is neutral only if it bases its recommendations on user interests—regardless of the content of the videos.

Later in the hearing, we hear from Justice Sotomayor:
\begin{quote}
[Justice Sotomayor:] I can really see that an internet provider who was in cahoots with ISIS provided them with an algorithm that would take anybody in the world and find them for them and—and do recruiting of people by showing them other videos that will lead them to ISIS.… \ I gave you an example earlier of an internet provider working directly with ISIS and doing an algorithm that—teaching them how to do an algorithm that will look for everybody who is just ISIS-related. \emph{There's more a collusion in the creation than a neutral algorithm.} (emphasis mine)
\end{quote}

\noindent  Now we see the idea that if there is collusion between the search engine operator and ISIS—if the recommendations were geared towards people joining ISIS—then the search engine would not be neutral. Putting these pieces together from the Justices in my terms: the recommendation system is neutral only if values external to its aim—values other than those of the users' interests—play no role in how the system recommends videos.

We find a similar idea voiced by Facebook in response to accusations of political bias:

\begin{quote}
But…, Cathcart [director of product development for Facebook's ‘NewsFeed'] said, Facebook strives for neutrality. ‘We want this [the NewsFeed] to show you what you're most interested in… We're not interested in adding our point of view.' \\
\citep{newton2016}
\end{quote}

\noindent Implicit here is the aim-based view of neutrality. NewsFeed's aim is to show users what they're interested in. In turn, NewsFeed is neutral only if values external to that aim—such as Facebook's (political) point of view—play no role in how NewsFeed delivers its results.\footnote{\label{winner}The kind of neutrality at issue here—in Facebook's statement, the Supreme Court oral arguments, and my discussion throughout the paper—takes the aim of a system as given, and characterises neutrality in terms of that aim. There is another sense of neutrality, though, on which the aim itself may not be neutral: building algorithms with certain aims may end up advantaging some groups, ideologies, or ways of life over others—even without any intent to do so. For related discussion, see \S\ref{lessons-2}, \citep{winner1980} on technologies in general, and \citep{patten2011} on liberal neutrality.}

\subsection{Is algorithmic neutrality possible?}\label{general-q2}

Neutrality is impossible if the aim of a system underdetermines how the system delivers its results. In such cases, values external to the system's aim necessarily supplement that aim. External values may also override that aim, but it isn't necessary that they do.

We've seen that underdetermination arises for a system with a certain multidimensional aim—relevance. We'll find underdetermination with other multidimensional aims, too. Our hiring algorithm from \S\ref{general-q1} aims to classify candidates on the basis of qualification. Qualification is, presumably, a multidimensional concept, and so the system's aim underdetermines how to classify candidates; neutrality is impossible. The scope of the possibility concerns only external values supplementing qualification, since it's inevitable that they do; it's not inevitable that external values override qualification.

Underdetermination has sources other than multidimensionality, and I will canvass a few. Because underdetermination is pervasive, neutrality is impossible for many—if not most—algorithmic systems that affect our lives.

Underdetermination may arise if a system's aim involves an \emph{arbitrary threshold}.\footnote{\citet{johnson2023} makes a similar point.} Consider an algorithm for use in a foster care system. (Similar algorithms are in fact used—by, for example, the Department of Human Services in Allegh\-eny, Pennsylvania \citep{allegheny2022}.) The algorithm, imagine, is used to identify whether it's safe for a child in foster care to return to their original family. The aim of the system is to categorise children as at a low, medium, or high risk of being abused if they were to return. How likely must abuse be for a child to be categorised as at high risk? 10\%? 20\%? 21\%? 50\%? In other words, what is the threshold of high risk, and similarly for low and medium risk?  The aim of categorising children at low, medium, or high risk itself underdetermines what these thresholds are, and in turn, underdetermines how to categorise children.

Underdetermination may arise if the system has a \emph{conjunctive aim}.\footnote{\citet{fazelpourdanks2021} make a similar point.} Imagine an algorithm for use in pre-trial detention decisions in the US judicial system, along the lines of those that are in fact widely used \citep{angwinetal2016}. Such decisions are supposed to be based on two factors: if the defendant is released, whether they will commit a crime (likelihood of recidivism) and whether they will fail to appear for a future court appearance (likelihood of flight). Based on \emph{both} of these factors, our algorithm assigns defendants a single score that represents their aptness for pre-trial detention. Imagine that one defendant has slightly higher risk of recidivism than another while having a slightly lower risk of flight. Deciding which defendant should receive a higher risk score (or if they should have the same score) is a matter of weighting the likelihood of recidivism against the likelihood of flight. To do so, we cannot appeal to the dual aims of predicting recidivism risk and predicting flight risk, since these two aims ‘disagree' with one another. The aims of predicting recidivism risk and flight risk therefore underdetermine how to assign the single risk score.

Underdetermination may have even more sources. \citet{dotan2020} and \citet{johnson2023}, among others, show that algorithmic systems are analogous to certain scientific practices. After all, algorithmic systems often aim to get at the truth (like predicting whether someone will commit a crime). Some argue that the aim of truth underdetermines how to conduct certain scientific practices (as I noted in \S\ref{q2}). If such arguments are sound, then so too will be analogous arguments that the aim of truth in certain algorithmic systems underdetermines how those systems deliver their results. 

\subsection{What is the normative significance of algorithmic neutrality?}\label{general-q3}

Earlier, I gave an argument, that when generalised, shows that algorithmic neutrality isn't  \emph{in and of itself} either good or bad, fair or unfair—and likewise for non-neutrality. Rather, the normative significance of (non)neutrality in a given algorithmic system hinges on which external values play a role in how the system delivers its results, and which roles those values play. 

Consider how values of social justice might play a role in our hiring algorithm. These external values could override the system's aim (qualification), ranking applicants from historically marginalised groups higher than more qualified candidates from other groups. What is the normative status of this situation? This is a question about a certain kind of affirmative action: whether values of social justice should override the aim of qualification. Put another way: it turns on the extent to which the aim of qualification is worth pursuing at the cost of social justice.

Values of social justice might also supplement qualification: when candidates from historically marginalised groups are incomparable (or tied) in qualification with other candidates, the algorithm ranks those from marginalised groups higher. What is the normative significance of this situation? This is a question about a different kind of affirmative action—in particular, positive action (in the sense discussed in \S\ref{lessons-1}). Put another way, it's a question of whether certain external values (those of social justice) should play a certain role (supplementing qualification) in how the system ranks applicants.

In \S\ref{lessons-2}, I distinguished two kinds of neutrality: characterised generally, they are \emph{global neutrality}, where external values of any kind play no role in how the system ranks pages, and \emph{local neutrality}, where external values of a certain kind play no role in how the system ranks pages. For example, our hiring algorithm might be locally neutral with respect to values concerning social categories, if values that concern them—whether those are values of social justice or racist, sexist, or ableist (etc.) values—play no role in how the hiring system ranks candidates. However, a system that is locally neutral with respect to certain values within a certain domain may nonetheless be susceptible to complaints of favouritism within that domain. For example, because qualification is multidimensional, the dimensions of qualification must be weighted some way rather than others in the hiring algorithm. Even if values in the domain of social categories play no role in this weighting, members of certain social groups may nonetheless be disadvantaged by one weighting versus another. In that case, they have a certain standing to complain: the system could just as well have weighted the dimensions of qualification in a way that does not disadvantage them.\footnote{For related issues, see discussion of \emph{target variable definition} in, for example, \citep{passibarocas2019, galaetal2025}.}

In \S\ref{lessons-3}, I distinguished two forms of bias, which, stated generally, are that an algorithmic system is \emph{overriding biased} if external values override the system's aim in how the system delivers its results; a system is \emph{supplementing biased} if values external to the system's aim supplement the system's aim in how the system delivers its results. Complaints of overriding bias are on firm conceptual footing, since—even in necessarily non-neutral algorithmic systems—it's not necessary that external values override the system's aim. Some complaints of supplementing bias are on firm footing; others aren't. Consider a necessarily non-neutral system: external values necessarily play a role in how the system delivers its results. It makes no sense to complain \emph{that} such a system is supplementing biased; that much is, by definition, inevitable. But there's no obstacle to complaining \emph{how} the system is supplementing biased—no obstacle to complaining about which external values supplement the system's aim.

With these normative considerations in place, we can return to what we saw in the introduction: it's common for Big Tech and technology industry in general to present algorithmic neutrality as an ideal and use it as a shield against criticism; Google and Facebook in particular hold neutrality as central to their self-conceptions. These moves are, as a matter of fact, sometimes in bad faith. But they also face a deeper problem. At a conceptual level, neutrality is a false ideal and faulty shield. 

Neutrality is a false ideal because it's not something to pursue for its own sake, as we've seen. If you aspire to neutrality itself, you blind yourself to the real normative questions about (non)neutrality: what values (if any) should play a role in how a system delivers its results, and what roles those values should play—which is often a matter of the extent to which the system's aim is worth pursuing. 

The fact that neutrality isn't worth pursuing for its own sake also makes neutrality a faulty shield. When accused of wrongdoing, it's a hollow defence to maintain that you're doing something that's not in and of itself right, good, or fair. Neutrality is also a faulty shield because it's often impossible. When accused of some wrong, it's a hollow defence to maintain that you're doing something that in fact you cannot do.

\newpage

\bibliography{mpb_references2}

\end{document}